\documentclass[10pt,conference,twoside]{IEEEtran}

\usepackage{cite}
\usepackage[T1]{fontenc}
\usepackage{graphicx}
\usepackage{amssymb}
\usepackage{amsmath}
\usepackage{paralist}
\usepackage{setspace}
\usepackage{microtype}
\usepackage{hyperref}
\usepackage{url}
\usepackage{balance}
\usepackage[caption=false,font=footnotesize]{subfig}
\usepackage{xcolor}
\usepackage{dsfont}

\newcommand{\figwidth}{0.83\columnwidth}

\newcommand{\quantize}{\mathcal{Q}}

\newcommand{\snr}{\rho}

\newcommand{\sindr}{\gamma}

\usepackage{amssymb}
\usepackage{amsfonts}
\usepackage{mathrsfs}
\usepackage{xspace}
\usepackage{bm}
\usepackage{upgreek}

\newcommand{\safemath}[2]{\newcommand{#1}{\ensuremath{#2}\xspace}}

\safemath{\bma}{\mathbf{a}}
\safemath{\bmb}{\mathbf{b}}
\safemath{\bmc}{\mathbf{c}}
\safemath{\bmd}{\mathbf{d}}
\safemath{\bme}{\mathbf{e}}
\safemath{\bmf}{\mathbf{f}}
\safemath{\bmg}{\mathbf{g}}
\safemath{\bmh}{\mathbf{h}}
\safemath{\bmi}{\mathbf{i}}
\safemath{\bmj}{\mathbf{j}}
\safemath{\bmk}{\mathbf{k}}
\safemath{\bml}{\mathbf{l}}
\safemath{\bmm}{\mathbf{m}}
\safemath{\bmn}{\mathbf{n}}
\safemath{\bmo}{\mathbf{o}}
\safemath{\bmp}{\mathbf{p}}
\safemath{\bmq}{\mathbf{q}}
\safemath{\bmr}{\mathbf{r}}
\safemath{\bms}{\mathbf{s}}
\safemath{\bmt}{\mathbf{t}}
\safemath{\bmu}{\mathbf{u}}
\safemath{\bmv}{\mathbf{v}}
\safemath{\bmw}{\mathbf{w}}
\safemath{\bmx}{\mathbf{x}}
\safemath{\bmy}{\mathbf{y}}
\safemath{\bmz}{\mathbf{z}}
\safemath{\bmzero}{\mathbf{0}}
\safemath{\bmone}{\mathbf{1}}

\bmdefine{\biad}{a}
\bmdefine{\bibd}{b}
\bmdefine{\bicd}{c}
\bmdefine{\bidd}{d}
\bmdefine{\bied}{e}
\bmdefine{\bifd}{f}
\bmdefine{\bigd}{g}
\bmdefine{\bihd}{h}
\bmdefine{\biid}{i}
\bmdefine{\bijd}{j}
\bmdefine{\bikd}{k}
\bmdefine{\bild}{l}
\bmdefine{\bimd}{m}
\bmdefine{\bind}{n}
\bmdefine{\biod}{o}
\bmdefine{\bipd}{p}
\bmdefine{\biqd}{q}
\bmdefine{\bird}{r}
\bmdefine{\bisd}{s}
\bmdefine{\bitd}{t}
\bmdefine{\biud}{u}
\bmdefine{\bivd}{v}
\bmdefine{\biwd}{w}
\bmdefine{\bixd}{x}
\bmdefine{\biyd}{y}
\bmdefine{\bizd}{z}

\bmdefine{\bixid}{\xi}
\bmdefine{\bilambdad}{\lambda}
\bmdefine{\bimud}{\mu}
\bmdefine{\bithetad}{\theta}
\bmdefine{\biphid}{\phi}
\bmdefine{\bideltad}{\delta}

\safemath{\bmia}{\biad}
\safemath{\bmib}{\bibd}
\safemath{\bmic}{\bicd}
\safemath{\bmid}{\bidd}
\safemath{\bmie}{\bied}
\safemath{\bmif}{\bifd}
\safemath{\bmig}{\bigd}
\safemath{\bmih}{\bihd}
\safemath{\bmii}{\biid}
\safemath{\bmij}{\bijd}
\safemath{\bmik}{\bikd}
\safemath{\bmil}{\bild}
\safemath{\bmim}{\bimd}
\safemath{\bmin}{\bind}
\safemath{\bmio}{\biod}
\safemath{\bmip}{\bipd}
\safemath{\bmiq}{\biqd}
\safemath{\bmir}{\bird}
\safemath{\bmis}{\bisd}
\safemath{\bmit}{\bitd}
\safemath{\bmiu}{\biud}
\safemath{\bmiv}{\bivd}
\safemath{\bmiw}{\biwd}
\safemath{\bmix}{\bixd}
\safemath{\bmiy}{\biyd}
\safemath{\bmiz}{\bizd}

\safemath{\bmxi}{\bixid}
\safemath{\bmlambda}{\bilambdad}
\safemath{\bmmu}{\bimud}
\safemath{\bmtheta}{\bithetad}
\safemath{\bmphi}{\biphid}
\safemath{\bmdelta}{\bideltad}

\safemath{\bA}{\mathbf{A}}
\safemath{\bB}{\mathbf{B}}
\safemath{\bC}{\mathbf{C}}
\safemath{\bD}{\mathbf{D}}
\safemath{\bE}{\mathbf{E}}
\safemath{\bF}{\mathbf{F}}
\safemath{\bG}{\mathbf{G}}
\safemath{\bH}{\mathbf{H}}
\safemath{\bI}{\mathbf{I}}
\safemath{\bJ}{\mathbf{J}}
\safemath{\bK}{\mathbf{K}}
\safemath{\bL}{\mathbf{L}}
\safemath{\bM}{\mathbf{M}}
\safemath{\bN}{\mathbf{N}}
\safemath{\bO}{\mathbf{O}}
\safemath{\bP}{\mathbf{P}}
\safemath{\bQ}{\mathbf{Q}}
\safemath{\bR}{\mathbf{R}}
\safemath{\bS}{\mathbf{S}}
\safemath{\bT}{\mathbf{T}}
\safemath{\bU}{\mathbf{U}}
\safemath{\bV}{\mathbf{V}}
\safemath{\bW}{\mathbf{W}}
\safemath{\bX}{\mathbf{X}}
\safemath{\bY}{\mathbf{Y}}
\safemath{\bZ}{\mathbf{Z}}

\safemath{\bZero}{\mathbf{0}}
\safemath{\bOne}{\mathbf{1}}
\safemath{\bDelta}{\mathbf{\Delta}}
\safemath{\bLambda}{\mathbf{\UpLambda}}
\safemath{\bPhi}{\mathbf{\Upphi}}
\safemath{\bSigma}{\mathbf{\Upsigma}}
\safemath{\bOmega}{\mathbf{\Upomega}}
\safemath{\bTheta}{\mathbf{\Uptheta}}

\bmdefine{\biAd}{A}
\bmdefine{\biBd}{B}
\bmdefine{\biCd}{C}
\bmdefine{\biDd}{D}
\bmdefine{\biEd}{E}
\bmdefine{\biFd}{F}
\bmdefine{\biGd}{G}
\bmdefine{\biHd}{H}
\bmdefine{\biId}{I}
\bmdefine{\biJd}{J}
\bmdefine{\biKd}{K}
\bmdefine{\biLd}{L}
\bmdefine{\biMd}{M}
\bmdefine{\biOd}{N}
\bmdefine{\biPd}{O}
\bmdefine{\biQd}{P}
\bmdefine{\biRd}{R}
\bmdefine{\biSd}{S}
\bmdefine{\biTd}{T}
\bmdefine{\biUd}{U}
\bmdefine{\biVd}{V}
\bmdefine{\biWd}{W}
\bmdefine{\biXd}{X}
\bmdefine{\biYd}{Y}
\bmdefine{\biZd}{Z}

\bmdefine{\biDelta}{\Delta}
\bmdefine{\biLambda}{\Lambda}
\bmdefine{\biPhi}{\Phi}
\bmdefine{\biSigma}{\Sigma}
\bmdefine{\biOmega}{\Omega}
\bmdefine{\biTheta}{\Theta}

\safemath{\bimA}{\biAd}
\safemath{\bimB}{\biBd}
\safemath{\bimC}{\biCd}
\safemath{\bimD}{\biDd}
\safemath{\bimE}{\biEd}
\safemath{\bimF}{\biFd}
\safemath{\bimG}{\biGd}
\safemath{\bimH}{\biHd}
\safemath{\bimI}{\biId}
\safemath{\bimJ}{\biJd}
\safemath{\bimK}{\biKd}
\safemath{\bimL}{\biLd}
\safemath{\bimM}{\biMd}
\safemath{\bimN}{\biNd}
\safemath{\bimO}{\biOd}
\safemath{\bimP}{\biPd}
\safemath{\bimQ}{\biQd}
\safemath{\bimR}{\biRd}
\safemath{\bimS}{\biSd}
\safemath{\bimT}{\biTd}
\safemath{\bimU}{\biUd}
\safemath{\bimV}{\biVd}
\safemath{\bimW}{\biWd}
\safemath{\bimX}{\biXd}
\safemath{\bimY}{\biYd}
\safemath{\bimZ}{\biZd}

\safemath{\bimDelta}{\biDelta}
\safemath{\bimLambda}{\biLambda}
\safemath{\bimPhi}{\biPhi}
\safemath{\bimSigma}{\biSigma}
\safemath{\bimOmega}{\biOmega}
\safemath{\bimTheta}{\biTheta}

\safemath{\setA}{\mathcal{A}}
\safemath{\setB}{\mathcal{B}}
\safemath{\setC}{\mathcal{C}}
\safemath{\setD}{\mathcal{D}}
\safemath{\setE}{\mathcal{E}}
\safemath{\setF}{\mathcal{F}}
\safemath{\setG}{\mathcal{G}}
\safemath{\setH}{\mathcal{H}}
\safemath{\setI}{\mathcal{I}}
\safemath{\setJ}{\mathcal{J}}
\safemath{\setK}{\mathcal{K}}
\safemath{\setL}{\mathcal{L}}
\safemath{\setM}{\mathcal{M}}
\safemath{\setN}{\mathcal{N}}
\safemath{\setO}{\mathcal{O}}
\safemath{\setP}{\mathcal{P}}
\safemath{\setQ}{\mathcal{Q}}
\safemath{\setR}{\mathcal{R}}
\safemath{\setS}{\mathcal{S}}
\safemath{\setT}{\mathcal{T}}
\safemath{\setU}{\mathcal{U}}
\safemath{\setV}{\mathcal{V}}
\safemath{\setW}{\mathcal{W}}
\safemath{\setX}{\mathcal{X}}
\safemath{\setY}{\mathcal{Y}}
\safemath{\setZ}{\mathcal{Z}}
\safemath{\emptySet}{\varnothing}

\safemath{\colA}{\mathscr{A}}
\safemath{\colB}{\mathscr{B}}
\safemath{\colC}{\mathscr{C}}
\safemath{\colD}{\mathscr{D}}
\safemath{\colE}{\mathscr{E}}
\safemath{\colF}{\mathscr{F}}
\safemath{\colG}{\mathscr{G}}
\safemath{\colH}{\mathscr{H}}
\safemath{\colI}{\mathscr{I}}
\safemath{\colJ}{\mathscr{J}}
\safemath{\colK}{\mathscr{K}}
\safemath{\colL}{\mathscr{L}}
\safemath{\colM}{\mathscr{M}}
\safemath{\colN}{\mathscr{N}}
\safemath{\colO}{\mathscr{O}}
\safemath{\colP}{\mathscr{P}}
\safemath{\colQ}{\mathscr{Q}}
\safemath{\colR}{\mathscr{R}}
\safemath{\colS}{\mathscr{S}}
\safemath{\colT}{\mathscr{T}}
\safemath{\colU}{\mathscr{U}}
\safemath{\colV}{\mathscr{V}}
\safemath{\colW}{\mathscr{W}}
\safemath{\colX}{\mathscr{X}}
\safemath{\colY}{\mathscr{Y}}
\safemath{\colZ}{\mathscr{Z}}

\safemath{\opA}{\mathbb{A}}
\safemath{\opB}{\mathbb{B}}
\safemath{\opC}{\mathbb{C}}
\safemath{\opD}{\mathbb{D}}
\safemath{\opE}{\mathbb{E}}
\safemath{\opF}{\mathbb{F}}
\safemath{\opG}{\mathbb{G}}
\safemath{\opH}{\mathbb{H}}
\safemath{\opI}{\mathbb{I}}
\safemath{\opJ}{\mathbb{J}}
\safemath{\opK}{\mathbb{K}}
\safemath{\opL}{\mathbb{L}}
\safemath{\opM}{\mathbb{M}}
\safemath{\opN}{\mathbb{N}}
\safemath{\opO}{\mathbb{O}}
\safemath{\opP}{\mathbb{P}}
\safemath{\opQ}{\mathbb{Q}}
\safemath{\opR}{\mathbb{R}}
\safemath{\opS}{\mathbb{S}}
\safemath{\opT}{\mathbb{T}}
\safemath{\opU}{\mathbb{U}}
\safemath{\opV}{\mathbb{V}}
\safemath{\opW}{\mathbb{W}}
\safemath{\opX}{\mathbb{X}}
\safemath{\opY}{\mathbb{Y}}
\safemath{\opZ}{\mathbb{Z}}
\safemath{\opZero}{\mathbb{O}}
\safemath{\identityop}{\opI}

\safemath{\veca}{\bma}
\safemath{\vecb}{\bmb}
\safemath{\vecc}{\bmc}
\safemath{\vecd}{\bmd}
\safemath{\vece}{\bme}
\safemath{\vecf}{\bmf}
\safemath{\vecg}{\bmg}
\safemath{\vech}{\bmh}
\safemath{\veci}{\bmi}
\safemath{\vecj}{\bmj}
\safemath{\veck}{\bmk}
\safemath{\vecl}{\bml}
\safemath{\vecm}{\bmm}
\safemath{\vecn}{\bmn}
\safemath{\veco}{\bmo}
\safemath{\vecp}{\bmp}
\safemath{\vecq}{\bmq}
\safemath{\vecr}{\bmr}
\safemath{\vecs}{\bms}
\safemath{\vect}{\bmt}
\safemath{\vecu}{\bmu}
\safemath{\vecv}{\bmv}
\safemath{\vecw}{\bmw}
\safemath{\vecx}{\bmx}
\safemath{\vecy}{\bmy}
\safemath{\vecz}{\bmz}

\safemath{\veczero}{\bmzero}
\safemath{\vecone}{\bmone}
\safemath{\vecxi}{\bmxi}
\safemath{\veclambda}{\bmlambda}
\safemath{\vecmu}{\bmmu}
\safemath{\vectheta}{\bmtheta}
\safemath{\vecphi}{\bmphi}
\safemath{\vecdelta}{\bmdelta}

\safemath{\matA}{\bA}
\safemath{\matB}{\bB}
\safemath{\matC}{\bC}
\safemath{\matD}{\bD}
\safemath{\matE}{\bE}
\safemath{\matF}{\bF}
\safemath{\matG}{\bG}
\safemath{\matH}{\bH}
\safemath{\matI}{\bI}
\safemath{\matJ}{\bJ}
\safemath{\matK}{\bK}
\safemath{\matL}{\bL}
\safemath{\matM}{\bM}
\safemath{\matN}{\bN}
\safemath{\matO}{\bO}
\safemath{\matP}{\bP}
\safemath{\matQ}{\bQ}
\safemath{\matR}{\bR}
\safemath{\matS}{\bS}
\safemath{\matT}{\bT}
\safemath{\matU}{\bU}
\safemath{\matV}{\bV}
\safemath{\matW}{\bW}
\safemath{\matX}{\bX}
\safemath{\matY}{\bY}
\safemath{\matZ}{\bZ}
\safemath{\matzero}{\bmzero}

\safemath{\matDelta}{\bDelta}
\safemath{\matLambda}{\bLambda}
\safemath{\matPhi}{\bPhi}
\safemath{\matSigma}{\bSigma}
\safemath{\matOmega}{\bOmega}
\safemath{\matTheta}{\bTheta}

\safemath{\matidentity}{\matI}
\safemath{\matone}{\matO}

\safemath{\rnda}{A}
\safemath{\rndb}{B}
\safemath{\rndc}{C}
\safemath{\rndd}{D}
\safemath{\rnde}{E}
\safemath{\rndf}{F}
\safemath{\rndg}{G}
\safemath{\rndh}{H}
\safemath{\rndi}{I}
\safemath{\rndj}{J}
\safemath{\rndk}{K}
\safemath{\rndl}{L}
\safemath{\rndm}{M}
\safemath{\rndn}{N}
\safemath{\rndo}{O}
\safemath{\rndp}{P}
\safemath{\rndq}{Q}
\safemath{\rndr}{R}
\safemath{\rnds}{S}
\safemath{\rndt}{T}
\safemath{\rndu}{U}
\safemath{\rndv}{V}
\safemath{\rndw}{W}
\safemath{\rndx}{X}
\safemath{\rndy}{Y}
\safemath{\rndz}{Z}

\safemath{\rveca}{\bimA}
\safemath{\rvecb}{\bimB}
\safemath{\rvecc}{\bimC}
\safemath{\rvecd}{\bimD}
\safemath{\rvece}{\bimE}
\safemath{\rvecf}{\bimF}
\safemath{\rvecg}{\bimG}
\safemath{\rvech}{\bimH}
\safemath{\rveci}{\bimI}
\safemath{\rvecj}{\bimJ}
\safemath{\rveck}{\bimK}
\safemath{\rvecl}{\bimL}
\safemath{\rvecm}{\bimM}
\safemath{\rvecn}{\bimN}
\safemath{\rveco}{\bomO}
\safemath{\rvecp}{\bimP}
\safemath{\rvecq}{\bimQ}
\safemath{\rvecr}{\bimR}
\safemath{\rvecs}{\bimS}
\safemath{\rvect}{\bimT}
\safemath{\rvecu}{\bimU}
\safemath{\rvecv}{\bimV}
\safemath{\rvecw}{\bimW}
\safemath{\rvecx}{\bimX}
\safemath{\rvecy}{\bimY}
\safemath{\rvecz}{\bimZ}

\safemath{\rvecxi}{\bmxi}
\safemath{\rveclambda}{\bmlambda}
\safemath{\rvecmu}{\bmmu}
\safemath{\rvectheta}{\bmtheta}
\safemath{\rvecphi}{\bmphi}

\safemath{\rmatA}{\bimA}
\safemath{\rmatB}{\bimB}
\safemath{\rmatC}{\bimC}
\safemath{\rmatD}{\bimD}
\safemath{\rmatE}{\bimE}
\safemath{\rmatF}{\bimF}
\safemath{\rmatG}{\bimG}
\safemath{\rmatH}{\bimH}
\safemath{\rmatI}{\bimI}
\safemath{\rmatJ}{\bimJ}
\safemath{\rmatK}{\bimK}
\safemath{\rmatL}{\bimL}
\safemath{\rmatM}{\bimM}
\safemath{\rmatN}{\bimN}
\safemath{\rmatO}{\bimO}
\safemath{\rmatP}{\bimP}
\safemath{\rmatQ}{\bimQ}
\safemath{\rmatR}{\bimR}
\safemath{\rmatS}{\bimS}
\safemath{\rmatT}{\bimT}
\safemath{\rmatU}{\bimU}
\safemath{\rmatV}{\bimV}
\safemath{\rmatW}{\bimW}
\safemath{\rmatX}{\bimX}
\safemath{\rmatY}{\bimY}
\safemath{\rmatZ}{\bimZ}

\safemath{\rmatDelta}{\bimDelta}
\safemath{\rmatLambda}{\bimLambda}
\safemath{\rmatPhi}{\bimPhi}
\safemath{\rmatSigma}{\bimSigma}
\safemath{\rmatOmega}{\bimOmega}
\safemath{\rmatTheta}{\bimTheta}

\usepackage{amssymb}
\usepackage{amsfonts}
\usepackage{mathrsfs}
\usepackage{xspace}
\usepackage{bm}
\usepackage{fancyref}
\usepackage{textcomp}

\usepackage{multirow}
\usepackage{stmaryrd}

\newenvironment{textbmatrix}{	\setlength{\arraycolsep}{2.5pt}%
								\big[\begin{matrix}}{\end{matrix}\big]%
								\raisebox{0.08ex}{\vphantom{M}}}

\def\be{\begin{equation}}
\def\ee{\end{equation}}
\def\een{\nonumber \end{equation}}
\def\mat{\begin{bmatrix}}
\def\emat{\end{bmatrix}}
\def\btm{\begin{textbmatrix}}
\def\etm{\end{textbmatrix}}

\def\ba#1\ea{\begin{align}#1\end{align}}
\def\bas#1\eas{\begin{align*}#1\end{align*}}
\def\bs#1\es{\begin{split}#1\end{split}} 
\def\bg#1\eg{\begin{gather}#1\end{gather}}
\def\bml#1\eml{\begin{multline}#1\end{multline}}
\def\bi#1\ei{\begin{itemize}#1\end{itemize}}

\newcommand{\subtext}[1]{\text{\fontfamily{cmr}\fontshape{n}\fontseries{m}\selectfont{}#1}}
\newcommand{\lefto}{\mathopen{}\left}

\DeclareMathOperator{\sign}{sgn}			
\DeclareMathOperator{\Exop}{\opE}			
\DeclareMathOperator{\Varop}{\mathrm{Var}}
\DeclareMathOperator{\Covop}{\mathrm{Cov}}
\DeclareMathOperator{\rot}{\nabla\times}		

\newcommand{\Ex}[2]{\ensuremath{\Exop_{#1}\lefto[#2\right]}} 	
\newcommand{\abs}[1]{\lefto\lvert#1\right\rvert}		

\newcommand{\vecnorm}[1]{\lefto\lVert#1\right\rVert}		

\safemath{\dirac}{\delta}					
\safemath{\krond}{\dirac}					

\safemath{\upto}{\uparrow}
\safemath{\downto}{\downarrow}
\safemath{\iu}{j}							
\safemath{\ev}{\lambda}						
\safemath{\hilseqspace}{l^{2}}				
\newcommand{\banachfunspace}[1]{\setL^{#1}}	
\safemath{\hilfunspace}{\banachfunspace{2}}	

\safemath{\SNR}{\textsf{SNR}} 				
\safemath{\PAR}{\textsf{PAR}} 				
\safemath{\No}{N_0}							
\safemath{\Es}{E_s}							
\safemath{\Eb}{E_b}							
\safemath{\EbNo}{\frac{\Eb}{\No}}
\safemath{\EsNo}{\frac{\Es}{\No}}

\DeclareMathOperator{\CHop}{\ensuremath{\opH}} 
\safemath{\tvir}{\rndh_{\CHop}}				
\safemath{\tvtf}{\rndl_{\CHop}}				
\safemath{\spf}{\rnds_{\CHop}}				
\safemath{\bff}{H_{\CHop}}					

\safemath{\ircf}{r_{h}}						
\safemath{\tftvcf}{r_{s}}					
\safemath{\tfcf}{r_{l}}						
\safemath{\bfcf}{r_{H}}						

\safemath{\tcorr}{c_h}						
\safemath{\scf}{c_{s}}						
\safemath{\tfcorr}{c_{l}}					
\safemath{\fcorr}{c_{H}}						

\safemath{\mi}{I}							
\safemath{\capacity}{C}						

\safemath{\normal}{\mathcal{N}}			
\safemath{\jpg}{\mathcal{CN}}			
\safemath{\mchain}{\leftrightarrow}		
\newcommand{\AIC}{\ensuremath{\text{AIC}}} 

\safemath{\dB}{\,\mathrm{dB}}
\safemath{\dBm}{\,\mathrm{dBm}}
\safemath{\Hz}{\,\mathrm{Hz}}
\safemath{\kHz}{\,\mathrm{kHz}}
\safemath{\MHz}{\,\mathrm{MHz}}
\safemath{\GHz}{\,\mathrm{GHz}}
\safemath{\s}{\,\mathrm{s}}
\safemath{\ms}{\,\mathrm{ms}}
\safemath{\mus}{\,\mathrm{\text{\textmu}s}}
\safemath{\ns}{\,\mathrm{ns}}
\safemath{\ps}{\,\mathrm{ps}}
\safemath{\meter}{\,\mathrm{m}}
\safemath{\mm}{\,\mathrm{mm}}
\safemath{\cm}{\,\mathrm{cm}}
\safemath{\m}{\,\mathrm{m}}
\safemath{\W}{\,\mathrm{W}}
\safemath{\mW}{\, \mathrm{mW}}
\safemath{\J}{\,\mathrm{J}}
\safemath{\K}{\,\mathrm{K}}
\safemath{\bit}{\,\mathrm{bit}}
\safemath{\nat}{\,\mathrm{nat}}

\safemath{\define}{\triangleq}			

\safemath{\equivalent}{\sim}
\safemath{\distas}{\sim}					
\safemath{\sdiff}{\Delta}				

\safemath{\reals}{\mathbb{R}}
\safemath{\positivereals}{\reals_{+}}
\safemath{\integers}{\mathbb{Z}}
\safemath{\posint}{\integers_{+}}
\safemath{\naturals}{\mathbb{N}}
\safemath{\posnaturals}{\naturals_{+}}
\safemath{\complexset}{\mathbb{C}}
\safemath{\rationals}{\mathbb{Q}}

\newcommand*{\fancyrefapplabelprefix}{app}		
\newcommand*{\fancyrefthmlabelprefix}{thm}		
\newcommand*{\fancyreflemlabelprefix}{lem}		
\newcommand*{\fancyrefcorlabelprefix}{cor}		
\newcommand*{\fancyrefdeflabelprefix}{def}		
\newcommand*{\fancyrefproplabelprefix}{prop}	
\newcommand*{\fancyrefobslabelprefix}{obs}		
\newcommand*{\fancyrefalglabelprefix}{alg}		
\newcommand*{\fancyrefasmlabelprefix}{asm}	    

\frefformat{vario}{\fancyrefseclabelprefix}{Section~#1}
\frefformat{vario}{\fancyrefthmlabelprefix}{Theorem~#1}
\frefformat{vario}{\fancyreflemlabelprefix}{Lemma~#1}
\frefformat{vario}{\fancyrefcorlabelprefix}{Corollary~#1}
\frefformat{vario}{\fancyrefdeflabelprefix}{Definition~#1}
\frefformat{vario}{\fancyrefobslabelprefix}{Observation~#1}
\frefformat{vario}{\fancyrefasmlabelprefix}{Assumption~#1}
\frefformat{vario}{\fancyreffiglabelprefix}{Fig.~#1}
\frefformat{vario}{\fancyrefapplabelprefix}{Appendix~#1} 
\frefformat{vario}{\fancyrefproplabelprefix}{Proposition~#1}
\frefformat{vario}{\fancyrefalglabelprefix}{Algorithm~#1}
\frefformat{vario}{\fancyrefeqlabelprefix}{(#1)}

\newcommand{\marginnote}[1]{\marginpar{\framebox{\framebox{#1}}}}
\safemath{\dictab}{[\,\dicta\,\,\dictb\,]}

\safemath{\ysig}{\bmy}
\safemath{\ysighat}{\hat{\ysig}}
\safemath{\ysigdim}{M}
\safemath{\xsig}{\bmx}
\safemath{\xsigdim}{N}
\safemath{\nx}{n_x}
\safemath{\zsig}{\bmz}
\safemath{\zsigdim}{\ysigdim}
\safemath{\rsig}{\bmr}
\safemath{\Adict}{\bA}
\safemath{\Adicttilde}{\widetilde{\Adict}}
\safemath{\Adictdim}{\outputdim\times\xsigdim}
\safemath{\avec}{\bma}
\safemath{\avectilde}{\tilde{\avec}}
\safemath{\Bdict}{\bB}
\safemath{\Bdicttilde}{\widetilde{\Bdict}}
\safemath{\Cdict}{\bC}
\safemath{\cvec}{\bmc}
\safemath{\Ddict}{\bD}
\safemath{\Ddictdim}{\ysigdim\times\xsigdim}
\safemath{\dvec}{\bmd}
\safemath{\Ddicttilde}{\widetilde{\bD}}
\safemath{\Bonb}{\bB}
\safemath{\bvec}{\bmb}
\safemath{\Bonbdim}{\ysigdim\times\ysigdim}
\safemath{\noise}{\bmn}
\safemath{\noisedim}{\ysigim}
\safemath{\err}{\bme}
\safemath{\errdim}{\ysigdim}
\safemath{\errset}{\setE}
\safemath{\nerr}{n_e}
\safemath{\delop}{\bP_\errset}
\safemath{\delopc}{\bP_{{\errset}^c}}

\safemath{\cplxi}{\imath}
\safemath{\cplxj}{\jmath}

\safemath{\dict}{\matD}
\safemath{\inputdim}{N}		
\safemath{\outputdim}{M}		
\safemath{\sparsity}{S}	
\safemath{\inputdimA}{{N_a}}	
\safemath{\inputdimB}{{N_b}}	
\safemath{\elemA}{{n_a}}	
\safemath{\elemB}{{n_b}}	
\safemath{\resA}{\matR_a}	
\safemath{\resB}{\matR_b}	
\safemath{\subD}{\matS} 
\safemath{\subA}{\matS_a} 
\safemath{\subB}{\matS_b} 
\safemath{\dicta}{\matA} 	
\safemath{\dictb}{\matB} 	
\safemath{\hollowS}{H}
\safemath{\hollowA}{H_a}
\safemath{\hollowB}{H_b}
\safemath{\cross}{Z}
\safemath{\coh}{\mu_d}			
\safemath{\coha}{\mu_a}			
\safemath{\cohb}{\mu_b}			
\safemath{\mubs}{\nu}	
\safemath{\cohm}{\mu_m} 
\safemath{\dictset}{\setD}	
\safemath{\dictsetp}{\dictset(\coh,\coha,\cohb)}	
\safemath{\dictsetgen}{\dictset_\text{gen}}
\safemath{\dictsetgenp}{\dictsetgen(\coh)}
\safemath{\dictsetonb}{\dictset_\text{onb}}
\safemath{\dictsetonbp}{\dictsetonb(\coh)}

\safemath{\leftside}{U}
\safemath{\rightsideA}{R_a}
\safemath{\rightsideB}{R_b}

\safemath{\indexS}{\setI_S} 

\safemath{\na}{n_a}			
\safemath{\nb}{n_b}			
\safemath{\coeffa}{p_i}	
\safemath{\coeffb}{q_j}	
\safemath{\seta}{\setP}		
\safemath{\setb}{\setQ}     
\safemath{\setw}{\setW}	
\safemath{\setz}{\setZ}	
\safemath{\cola}{\veca}		
\safemath{\colb}{\vecb}		
\safemath{\cold}{\vecd}		
\safemath{\inputvec}{\vecx} 	
\safemath{\error}{\vece}	
\safemath{\noiseout}{\vecz} 	
\safemath{\inputvecel}{x}
\safemath{\inputveca}{\vecx_a}
\safemath{\inputvecb}{\vecx_b}
\safemath{\outputvec}{\vecy}	
\safemath{\lambdamin}{\lambda_{\mathrm{min}}}

\safemath{\elltwo}{\ell_2}
\safemath{\ellone}{\ell_1}
\safemath{\ellzero}{\ell_0}
\safemath{\ellinf}{\ell_\infty}
\safemath{\ellinftilde}{\ell_{\widetilde\infty}}
\safemath{\licard}{Z(\coh,\coha,\cohb)}
\safemath{\xsol}{\hat{x}}
\safemath{\xbord}{x_b}		
\safemath{\xstat}{x_s}		
\safemath{\xstatLone}{\tilde{x}_s}
\safemath{\order}{\mathcal{O}} 
\safemath{\scales}{\Theta} 
\safemath{\ones}{\mathbf{1}} 
\safemath{\zeroes}{\mathbf{0}} 
\safemath{\thlone}{\kappa(\coh,\cohb)} 
\safemath{\constoneA}{\delta} 
\safemath{\constoneB}{\epsilon} 
\safemath{\nlarge}{L}				   
\safemath{\sumlarge}{S_\nlarge}
\safemath{\maxlarger}{P_\nlarge}	   
\safemath{\Pzero}{\textrm{P0}}	
\safemath{\Pone}{\textrm{P1}}
\safemath{\vecfir}{\vecw}			 
\safemath{\vecsec}{\vecz}
\safemath{\elvecfir}{w}              
\safemath{\elvecsec}{z}				 
\safemath{\nlargefir}{n}
\safemath{\normout}{\gamma}
\safemath{\auxfun}{h}
\safemath{\supp}{\textrm{supp}}

\safemath{\indexa}{\ell}
\safemath{\indexb}{r}
\safemath{\indexc}{i}
\safemath{\indexd}{j}

\safemath{\project}{P}

\IEEEoverridecommandlockouts

\begin{document}

%
\title{Massive MU-MIMO-OFDM Downlink with \\ One-Bit~DACs and Linear Precoding}
\author{\IEEEauthorblockN{Sven Jacobsson$^{1,2}$, Giuseppe Durisi$^{1}$, Mikael Coldrey$^{2}$, and Christoph Studer$^{3}$}
\thanks{The work of SJ and GD was partly supported by the Swedish Foundation for Strategic Research under grant ID14-0022, and by the Swedish Governmental Agency for Innovation Systems (VINNOVA) within the competence center ChaseOn. The work of CS was supported in part by Xilinx Inc. and by the US National Science Foundation (NSF) under grants ECCS-1408006, CCF-1535897, CAREER CCF-1652065, and CNS-1717559.}
\IEEEauthorblockA{
$^{1}$Chalmers University of Technology, Gothenburg, Sweden\\
$^{2}$Ericsson Research, Gothenburg, Sweden\\
$^{3}$Cornell University, Ithaca, NY, USA
}}
\maketitle

\begin{abstract}
Massive multiuser (MU) multiple-input multiple-output (MIMO) is foreseen to be a key technology in future wireless communication systems. 
In this paper, we analyze the downlink performance of an orthogonal frequency division multiplexing (OFDM)-based massive MU-MIMO system in which the base station~(BS) is equipped with 1-bit digital-to-analog converters (DACs).
Using Bussgang's theorem, we characterize the  performance achievable with linear precoders (such as maximal-ratio transmission and zero forcing) in terms of bit error rate (BER).
Our analysis accounts for the possibility of oversampling the time-domain transmit signal before the DACs.
We further develop a lower bound on the information-theoretic sum-rate throughput achievable with Gaussian inputs.

Our results suggest that the performance achievable with 1-bit DACs in a massive MU-MIMO-OFDM downlink are satisfactory provided that the number of BS antennas is sufficiently large. 

\end{abstract}

\section{Introduction}

Massive multiuser (MU) multiple-input multiple-output (MIMO) is foreseen to be one of the key technologies in fifth-generation~(5G) cellular systems. 
Massive MU-MIMO deploys a large number of antennas at the base station~(BS), which enables significant improvements in terms of spectral efficiency, energy efficiency, reliability, and coverage compared to traditional small-scale MIMO~\cite{rusek14a, larsson14a}. 

However, increasing the number of BS antennas by orders of magnitude may lead to a significant growth in the circuit power consumption, system costs, and hardware complexity. 
To compensate for this, it is necessary to deploy practical massive MU-MIMO systems with cheap, power-efficient, and low-complexity hardware components.
However, such inexpensive and energy-efficient analog circuitry may fundamentally limit the capacity of the system, due to hardware imperfections. 

One of the most dominant sources of power consumption in massive MU-MIMO architectures are the data converters at the BS. 
Today's state-of-the-art multi-antenna BSs deploy high-resolution (e.g., 10-bit or more) digital-to-analog converters~(DACs) at each radio frequency (RF) port for downlink transmission. 
However, scaling up this approach to massive MU-MIMO systems with hundreds or thousands of BS antenna elements seems infeasible. 
Indeed, apart from system costs and power consumption, such an approach may also result in a bandwidth overload on the fronthaul link connecting the baseband processing unit to the RF ports.
A natural approach to address all of these issues is to reduce the resolution of the DACs used at the BS.
In this paper, we shall consider the extreme case of a BS equipped with 1-bit-resolution DACs and investigate the downlink performance achievable with linear precoders over frequency-selective fading channels.

\begin{figure*}[t]
\centering
 \includegraphics[width=0.75\textwidth]{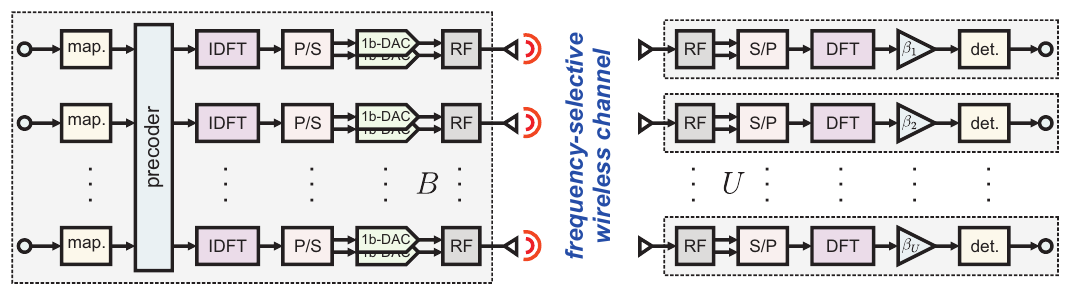}
 \vspace{-0.2cm}
 \caption{Overview of the proposed quantized massive MU-MIMO-OFDM downlink system. Left: a $B$-antenna BS that performs  linear precoding in the frequency-domain, transforms the precoded vector into time domain, and quantizes its entries using 1-bit DACs. Right: $U$ single-antenna UEs. } \label{fig:overview}
\vspace{-0.1cm}
\end{figure*}

\subsubsection*{Relevant prior art}
The performance achievable in the massive MU-MIMO \emph{uplink} (users communicate to the BS) when the BS is equipped with 1-bit analog-to-digital converters (ADCs) has been  investigated recently in~\cite{wen15b, mollen16c, jacobsson17b, li17b}. 
These results have shown that 1-bit-ADC massive MU-MIMO architectures yield high information-theoretic sum-rate throughput despite the severe nonlinearity introduced by the 1-bit ADCs, provided that the number of BS antennas is sufficiently large. 

In contrast to these results, far less is known about the impact on performance of the use of 1-bit~DACs in the massive~MU-MIMO~\emph{downlink} (BS communicates to users).
For the \emph{frequency-flat} case, the performance of linear precoders such as maximal-ratio transmission (MRT) and zero-forcing~(ZF) in the massive MU-MIMO downlink with 1-bit DACs at the BS has been analyzed in~\cite{saxena16b, li17a, jacobsson17d}.
Similarly to the uplink case, it has been shown that it is possible to achieve low bit error rates~(BERs) and high sum-rate throughput despite the nonlinearity introduced by the 1-bit DACs.
More sophisticated \emph{nonlinear} precoders were  proposed recently, again for the frequency-flat case, in~\cite{jacobsson17d, jacobsson16d, castaneda17a, jedda16a,  jedda16b, tirkkonen17a}. 
These  nonlinear precoders yield significant performance gains over linear precoders, at the cost of an increased signal-processing complexity.
It is, however, an open question whether the nonlinear precoders proposed in the literature can be extended to  frequency-selective channels while maintaining a low computational complexity.

All results reviewed so far deal with single-carrier modulation over a frequency-flat channel.
To our knowledge, the only work that considers 1-bit DACs combined with orthogonal frequency division multiplexing~(OFDM)---the setup we shall focus on in the present paper---is~\cite{guerreiro16a}.
 There, using an approximate model for the distortion caused by the 1-bit DACs at the BS, it is shown that  MRT precoding yields a manageable distortion level at the user equipments (UEs), provided that the number of BS antennas is sufficiently~large. 
 However, the analysis in~\cite{guerreiro16a} is limited to symbol-rate sampling DACs and the performance of this architecture in terms of BER or achievable rates is not discussed.

\subsubsection*{Contributions}
We characterize the performance achievable in the MU-MIMO-OFDM downlink for systems in which the BS is equipped with 1-bit DACs.
In contrast to previous works~\cite{saxena16b, li17a, jedda16a, jedda16b, jacobsson17d, jacobsson16d, castaneda17a}, we assume OFDM transmission over a frequency-selective channel. 
Furthermore, our analysis supports oversampling DACs. In contrast, the results available in the literature apply only to the case of symbol-rate sampling.
Using Bussgang's theorem~\cite{bussgang52a}, we develop a closed-form expression for the signal-to-interference-noise-and-distortion ratio (SINDR) of linear precoders. 
We then use this expression to obtain (i)  an accurate approximation of the uncoded BER achievable with QPSK signaling and (ii) a lower bound on the achievable sum-rate downlink throughput. This lower bound corresponds to the rate achievable using Gaussian signaling and mismatched nearest-neighbor decoding at the UEs. 
We use numerical simulations to demonstrate that the use of 1-bit DACs and linear precoders in a massive MU-MIMO-OFDM system operating over a frequency-selective Rayleigh-fading channel yields high downlink sum-rate throughput and~low~BER. 

\subsubsection*{Notation}
Lowercase and uppercase boldface letters designate column vectors and matrices, respectively. 
For a matrix~$\bA$, we denote its transpose and its Hermitian transpose by~$\bA^T$, and~$\bA^H$, respectively. The entry on the $k$th row and on the $\ell$th column of the matrix $\matA$ is denoted as $[\matA]_{k,\ell}$.
Furthermore, the $k$th entry of  a vector $\veca$  is denoted as~$[\veca]_k$.
%
%
The main diagonal of~$\bA$ is~$\text{diag}(\bA)$. The matrix $\text{diag}(\veca)$ is a diagonal matrix with the elements of $\veca$ along its main diagonal.
The $M\times M$ identity matrix and the $M\times N$ all-zeros matrix are denoted by $\bI_M$ and~$\mathbf{0}_{M\times N}$, respectively.
The real and imaginary parts of a complex-valued vector $\veca$ are $\Re\{\veca\}$ and $\Im\{\veca\}$, respectively. 
%
%
We use~$\text{sgn}(\cdot)$ to denote the signum function, which is applied entry-wise to vectors and is defined as~$\text{sgn}(a)=1$ if~$a\ge0$ and~$\text{sgn}(a)=-1$ if~$a<0$.
%
%
The multivariate complex-valued circularly-symmetric Gaussian probability density function with covariance matrix~$\bK$ is denoted by~$\setC\setN(\bZero,\bK)$. We use~$\Ex{x}{\cdot}$ to denote expectation with respect to the random variable~$x$ and $Q(x) = \frac{1}{\sqrt{2\pi}}\int_{x}^\infty \exp\lefto( -y^2/2 \right) \text{d}y$ to denote the tail probability of the standard normal distribution.

%


\section{System Model} \label{sec:system}

We consider the downlink of a single-cell massive MU-MIMO-OFDM as depicted in \fref{fig:overview}. 
The BS, which is equipped with $B$ antennas, serves $U$ single-antenna UEs simultaneously and in the same frequency band. 
At the BS, a precoder maps the data symbols into precoded symbols  to be sent to the RF ports. Since the system is assumed to operate over a frequency selective channel, OFDM is used. 
Specifically, the frequency-domain precoded vector is transformed into a time-domain vector by performing an inverse discrete Fourier transform~(IDFT) at each RF port. Then, the resulting signal is generated by a pair of 1-bit DACs, one operating on the real part and one on the imaginary part of the time-domain signal.
At the UEs, the received time-domain  signal is mapped back to the frequency domain through a DFT.

\subsection{OFDM Numerology}
Each OFDM symbol consists of $S$ data symbols and $N$ subcarriers. Let~$\Delta f$ be the subcarrier spacing and $f_s = N \Delta f$ the sampling rate of the DACs.
%
We use the disjoint sets $\setS_d$ and $\setS_g$ to denote the set of subcarriers designated for data symbols ({occupied subcarriers}) and for guard subcarriers, respectively. The number of occupied subcarriers is~$\abs{\setS_d} = S$, and the number of guard subcarriers is~$\abs{\setS_g} = N-S$. 
Let~$\vecs_k$~denote the $U$-dimensional data vector associated with the $k$th~($k=0,\dots, N-1$) subcarrier.
We shall assume that $\Ex{}{\vecs_k\vecs_k^H} = \matI_U$ if $k \in \setS_d$ and that $\vecs_k = \veczero_{U\times 1}$ if $k \in \setS_g$.
%
Note that the case $S=N$ corresponds to symbol-rate sampling whereas setting $S< N$ yields an oversampling ratio (OSR) of~$N/S > 1$.

\subsection{Time-Domain Channel Input-Output Relation}

For simplicity, we  assume that all RF hardware components other than the DACs (e.g., local oscillators, mixers, power amplifiers, etc.) are ideal and that the ADCs at the UEs have infinite resolution. 
We also assume that the sampling rate~$f_s$ of the DACs at the BS is equal to the sampling rate of the ADCs at the UEs and that the system is perfectly synchronized. 
Under these assumptions, the received discrete-time baseband vector $\vecy_n \in \complexset^U$ containing the samples at discrete time $n$ of the  signals received at the $U$ UEs  can be written as
\begin{IEEEeqnarray}{rCl} \label{eq:received_time_mimo}
\vecy_n &=& \sum_{\ell=0}^{L-1} \matH_\ell	 \vecx_{n-\ell} + \vecw_n, \quad n=0,\dots, N-1.
\end{IEEEeqnarray}
Here, $\vecx_n \in \complexset^B$ is the quantized time-domain precoded vector at time $n$, the matrix~$\matH_\ell \in \opC^{U \times B}$ is the time-domain channel matrix associated with the $\ell$th channel tap of the frequency-selective channel ($\ell=0,\dots,L-1$). 
This matrix has entries~$\lefto[ \matH_\ell \right]_{u,b} \distas \jpg(0,1/\sqrt{L})$. We further assume the entries of the matrices $\{\matH_\ell\}$  to be independent and to remain constant for the duration of an OFDM symbol. 
Note that these assumptions yield a spatially white frequency-selective Rayleigh-fading channel with uniform power-delay profile. 
An extension to spatially correlated channels and to other power-delay profiles is straightforward.
Finally, $\vecw_n \distas \jpg\lefto( \veczero_{U \times 1}, N_0 \matI_U \right)$ denotes the additive white Gaussian noise~(AWGN) at the UEs. 
Here,~$N_0$ stands for the noise power spectral density (PSD).

%

At time $n$, the quantized time-domain precoded vector $\vecx_n$ is obtained by passing the infinite-resolution time-domain precoded vector $\vecz_n \in \opC^B$  through a set of 1-bit DACs.\footnote{In \fref{sec:linear_precoding}, we shall describe how this vector is obtained from the data symbols~$\{\vecs_k\}$ for  ${k \in \setS_d}$.} 
Specifically, let $\setX = \sqrt{{PS}/(2BN)} \{ 1+j,-1+j,-1-j,1-j\}$, where $P$ stands for the average transmit power at the BS.
We have that $\vecx_n = \quantize\lefto(\vecz_n\right)$ where the nonlinear function $\setQ(\cdot): \opC^B \rightarrow \setX^B$, which captures the  operation of the $2B$ DACs, is defined as
\begin{IEEEeqnarray}{rCl} \label{eq:quantizer_1bit}
\quantize\lefto(\vecz_n\right) &=& \sqrt{\frac{PS}{2BN}} \lefto(\sign\lefto(\Re\lefto\{ \vecz_n \right\}\right) + j\sign\lefto(\Im\lefto\{ \vecz_n \right\}\right)\right).
\end{IEEEeqnarray}

A cyclic prefix of length $L-1$ is prepended to the time-domain precoded vector $\vecz_n$  and is later discarded at the receive side. 
We shall not explicitly prepend the cyclic prefix to $\vecz_n$ to keep notation compact.
The cyclic prefix makes the channel matrix circulant and, hence, diagonalizable through IDFT and DFT operations at the BS and UE sides. 

\subsection{Frequency-Domain Channel Input-Output Relation}

Let $\matX = \lefto[ \vecx_0, \, \ldots, \, \vecx_{N-1} \right] \in \setX^{B\times N}$ and $\matY = \lefto[ \vecy_0, \, \ldots, \, \vecy_{N-1} \right]\in \complexset^{U\times N}$ be the time-domain transmitted and received matrices over the~$N$ time instants, respectively. 
Furthermore, let $\hat\matX  = \matX\matF^T $ and $\hat\matY  = \matY\matF^T$ be the corresponding frequency-domain matrices. 
Here,~$\matF$ stands for the $N \times N$ DFT matrix, which satisfies $\matF\matF^H = \matI_N$.
Finally,~let
\begin{IEEEeqnarray}{rCl}
	\hat\matH_{k} &=& \sum_{\ell=0}^{L-1} \matH_\ell \exp\lefto( - jk\frac{2\pi}{N}\ell \right), \quad k=0,\dots, N-1 \IEEEeqnarraynumspace
\end{IEEEeqnarray}
be the $U \times B$ frequency-domain channel matrix associated with the $k$th subcarrier.
After discarding the cyclic prefix, we can write the frequency-domain input-output relation at the $k$th subcarrier~as
\begin{IEEEeqnarray}{rCl} \label{eq:received_freq_mimo}
	\hat\vecy_{k}	&=& \hat\matH_{k}\hat\vecx_{k} + \hat\vecw_{k}.
\end{IEEEeqnarray}
Here, $\hat\vecx_{k}$ and $\hat\vecy_{k}$ are the $k$th column of $\hat\matX$ and $\hat\matY$, respectively.
Furthermore,~$\hat\vecw_{k} \distas \jpg(\veczero_{U \times 1}, N_0 \matI_U)$ is  the $k$th column of the matrix $\hat\matW = \matW\matF^T$, where $\matW = \lefto[ \vecw_0, \ldots, \vecw_{N-1} \right]$.

\subsection{Linear Precoding}
\label{sec:linear_precoding}
In this paper, we shall focus exclusively on linear precoding strategies.
We assume that the BS has perfect knowledge of the realizations of the frequency-domain channel matrices $\{\hat\matH_{k}\}$ for~$k \in \setS_d$ (a relaxation to the case of imperfect channel-state information will be considered in future works).
We further assume that the time-domain precoded vector $\vecz_n$ is obtained from the data symbols $\{\vecs_k\}$ as~follows: 
\begin{IEEEeqnarray}{rCl} \label{eq:vecz_time}
\vecz_n &=& \frac{1}{\sqrt{N}}\sum_{k=0}^{N-1} \hat\matP_{k} \vecs_k \exp\lefto(jk\frac{2\pi}{N}n\right)
\end{IEEEeqnarray}
for $n = 0,\,\ldots\,, N-1$. 
In words, the data symbols on the $k$th subcarrier are multiplied with the frequency-domain \emph{precoding matrix}~$\hat\matP_{k} \in \opC^{B \times U}$. The resulting frequency domain vector is then converted into time domain through an IDFT. 
We use the convention that~$\hat\matP_{k} = \matzero_{B \times U}$ on all guard subcarriers, i.e., for all subcarriers that satisfy~$k \in \setS_g$. 

We focus on two linear precoders that are commonly used in the infinite-resolution case, namely MRT and ZF. 
With MRT, the BS maximizes the power directed towards each UE, ignoring MU interference.
This is done by setting the precoding matrix equal to the Hermitian transpose of the channel matrix. 
With ZF, the BS avoids MU interference by setting the precoding matrix equal to the pseudo-inverse of the channel matrix.
Mathematically, we have
\begin{IEEEeqnarray}{rCl} \label{eq:precoder}
	\hat\matP_{k} &=& 
\begin{cases}
 	\frac{1}{\beta_\text{MRT}B}\hat\matH_{k}^H, & \text{for MRT}\\
	\frac{1}{\beta_\text{ZF}}\hat\matH_{k}^H \lefto( \hat\matH_{k} \hat\matH_{k}^H \right)^{-1}, & \text{for ZF}.
\end{cases}
\end{IEEEeqnarray}
Here, $\beta_\text{MRT}$ and $\beta_\text{ZF}$ are scaling factors chosen to ensure that the power constraint  is satisfied in the infinite-resolution case.

\section{Performance Analysis} \label{sec:lowerbound}

In the infinite-resolution case, the frequency-domain received signal $\hat\vecy_{k}$ can be written as
\begin{IEEEeqnarray}{rCl} 
	\hat\vecy_{k}	&=& \hat\matH_{k}\hat\matP_{k}\vecs_{k} + \hat\vecw_{k}.
\end{IEEEeqnarray}
In words, the received signal on subcarrier $k \in \setS_d$ depends only on $\vecs_k$ and not on the data symbols transmitted on other subcarriers. Hence, each subcarrier can be analyzed separately.
In the 1-bit-DAC case, however, the received signal on one subcarrier depends, in general, on the data symbols transmitted on all other subcarriers.
For a performance analysis, it is convenient to use the Kronecker product property $\text{vec}\lefto(\matA\matB\matC\right) = \lefto( \matC^T \otimes \matA \right) \text{vec} \lefto(\matB\right)$ and to write the frequency-domain received signal $\hat\matY$ in vectorized form $\hat\vecy = \text{vec}(\hat\matY) \in \opC^{UN}$~as
\begin{IEEEeqnarray}{rCl} \label{eq:vecz_precoded_vectorized_temp} 
	\hat\vecy &=& \hat\matH \lefto( \matF \otimes \matI_B \right) \vecx + \hat\vecw.
\end{IEEEeqnarray}
Here, $\vecx =\text{vec} \lefto(\matX\right) \in \setX^{BN}$, $\hat\vecw = \text{vec}(\hat\matW) \in \opC^{UN}$, and  $\hat\matH$ is the $U N \times B N$ block-diagonal matrix that has the matrices $\hat\matH_{0},\,\ldots,\,\hat\matH_{N-1}$ on its main diagonal.
In~\eqref{eq:vecz_precoded_vectorized_temp}, the quantized time-domain precoded vector $\vecx$ is given by
%
	$\vecx = \quantize(\vecz)$
where~$\vecz = \text{vec}\lefto( \matZ \right) \in \opC^{BN}$ and where $\matZ = \lefto[ \vecz_0, \ldots, \vecz_{N-1}\right]$. 
Now let~$\hat\matP \in \opC^{B N \times U N}$ denote the block-diagonal matrix that has the matrices $\hat\matP_{0},\,\ldots,\,\hat\matP_{N-1}$ on its main diagonal. 
We can write the time-domain precoded vector~$\vecz$~as
\begin{IEEEeqnarray}{rCl} \label{eq:vecz_precoded_vectorized}
\vecz = \lefto( \matF^H \! \otimes \matI_B \right) \hat\matP \vecs
\end{IEEEeqnarray}
where $\vecs = \text{vec}(\matS)$, and where $\matS = \lefto[ \vecs_0, \ldots, \vecs_{N-1}\right]$. 
Now, using~\eqref{eq:vecz_precoded_vectorized} in~\eqref{eq:vecz_precoded_vectorized_temp},  we obtain
\begin{IEEEeqnarray}{rCl} \label{eq:received_freq_mimo_vectorized}
	\hat\vecy &=& \hat\matH \lefto( \matF \otimes \matI_B \right) \quantize\lefto( \lefto( \matF^H \! \otimes \matI_B \right) \hat\matP \vecs\right) + \hat\vecw.
\end{IEEEeqnarray}
Next, we use Bussgang's theorem~\cite{bussgang52a} to decompose~\eqref{eq:received_freq_mimo_vectorized} in a form that enables analytic analysis. 
Bussgang's theorem has previously been used in, e.g., \cite{li17b, jacobsson17b, saxena16b, li17a, jacobsson17d}, to characterize the performance of the quantized single-carrier massive MU-MIMO uplink and downlink on frequency-flat channels. Here, we generalize these analyses for the downlink to the frequency-selective case.

\subsection{Decomposition Using Bussgang's Theorem}

The 1-bit DACs introduce an error $\quantize(\vecz) - \vecz$ that is \emph{correlated} with $\vecz$. 
For Gaussian inputs, Bussgang's theorem~\cite{bussgang52a} allows us to decompose $\quantize(\vecz)$ into two components: a linear function of $\vecz$ and a distortion that is \emph{uncorrelated} with~$\vecz$.
Specifically, let $\vecs_k\distas\jpg( \veczero_{U \times 1}, \matI_{U})$ for all $k \in \setS_d$.
By using Bussgang's theorem, the quantized time-domain precoded vector~$\vecx$ can be written as \cite{rowe82a, jacobsson17d}
\begin{IEEEeqnarray}{rCl}  \label{eq:inout_distortion}
	\vecx &=& \quantize(\vecz) = \matG \vecz + \vecd
\end{IEEEeqnarray}
where $\vecd \in \opC^{BN}$ is a distortion that is uncorrelated with $\vecz$.
Furthermore, $\matG = \matI_N \otimes \text{diag}\lefto( \vecg \right) \in \opR^{BN \times BN}$ is a diagonal matrix, where~(cf.~\cite[Eq.~(14)]{jacobsson17d})
\begin{IEEEeqnarray}{rCl} \label{eq:gainmatrix_1bit}
\text{diag}(\vecg) &=& \sqrt{\frac{2PS}{\pi BN}	} \, \text{diag}\lefto( \frac{1}{N}  \sum_{k=0}^{N-1}  \hat\matP_{k} \hat\matP_{k}^H \right)^{-1/2}. \IEEEeqnarraynumspace
\end{IEEEeqnarray}
%
%
Inserting~\eqref{eq:inout_distortion} into \eqref{eq:received_freq_mimo_vectorized} we obtain
\begin{IEEEeqnarray}{rCl}
\hat\vecy 
&=& \hat\matH \lefto( \matF \otimes \matI_B \right) \lefto(\matG \lefto( \matF^H \! \otimes \matI_B \right) \hat\matP  \vecs +  \vecd\right) + \hat\vecw \IEEEeqnarraynumspace\\	
&=& \hat\matH \matG \hat\matP  \vecs + \hat\matH \lefto( \matF \otimes \matI_B \right) \vecd + \hat\vecw. \label{eq:received_final}
\end{IEEEeqnarray}
The last step follows because $\lefto( \matF \otimes \matI_B \right) \matG \lefto( \matF^H \! \otimes \matI_B \right) = \matG$.

\subsection{Achievable Sum-Rate with Gaussian Inputs}

Let $\hat{y}_{u,k} = \lefto[ \hat\vecy_{k} \right]_u$ denote the  received signal on the $k$th subcarrier at the $u$th UE. 
It follows from~\eqref{eq:received_final} that
\begin{IEEEeqnarray}{rCl} 
\hat{y}_{u,k} &=&	\lefto[\hat\matH_{k\,} \text{diag}\lefto(\vecg\right) \hat\matP_{k}\right]_{u,u} s_{u,k} \nonumber\\
&&+\,\textstyle{\sum_{v \neq u}} \lefto[\hat\matH_{k\,} \text{diag}\lefto(\vecg\right) \hat\matP_{k}\right]_{u,v} s_{v,k} \nonumber\\
&&+\,[\hat\matH \lefto( \matF \otimes \matI_B \right) \vecd]_ {u+kU} + \hat{w}_{u,k}.\label{eq:decomposition}
\end{IEEEeqnarray}
The first term on the right-hand side of~\eqref{eq:decomposition} corresponds to the desired signal; the second term captures the MU interference; the third term describes the distortion introduced by the quantizer; the fourth term represents AWGN.

Let now $\sindr_{u,k}(\hat\matH)$ be the SINDR on the $k$th subcarrier for the~$u$th~UE.
Using~\eqref{eq:decomposition}, we can express $\sindr_{u,k}(\hat\matH)$ as
\begin{IEEEeqnarray}{rCl} \label{eq:sindr_mimo}
\sindr_{u,k}(\hat\matH) &=& \frac{ \lefto[\big\lvert\hat\matH_{k\,} \text{diag}\lefto(\vecg\right) \hat\matP_{k}\big\rvert^2\right]_{u,u} }{ \sum\limits_{v \neq u}\lefto[\big\lvert\hat\matH_{k\,}\text{diag}\lefto( \vecg \right) \hat\matP_{k}\big\rvert^2\right]_{u,v} \!\! + D_{u,k}(\hat\matH) + N_0} \IEEEeqnarraynumspace
\end{IEEEeqnarray}
where we have defined
\begin{IEEEeqnarray}{rCl}
	D_{u,k}(\hat\matH) &=& \lefto[ \hat\matH \lefto( \matF \otimes \matI_B \right) \matC_{\vecd} \lefto( \matF^H \!\! \otimes \matI_B \right)\hat\matH^H\right]_{u + kU, \, u + kU} \IEEEeqnarraynumspace
\end{IEEEeqnarray}
where $\matC_\vecd = \Ex{}{\vecd\vecd^H} \in \opC^{BN \times BN}$. 
Let now 
\begin{IEEEeqnarray}{rCl}
  \matC_\vecz = \lefto( \matF^H \! \otimes \matI_B \right) \hat\matP \hat\matP^H \lefto( \matF \otimes \matI_B \right).
\end{IEEEeqnarray}
Since $\vecz$ and $\vecd$ are uncorrelated, it follows from~\eqref{eq:inout_distortion} that
\begin{IEEEeqnarray}{rCl} \label{eq:Cdd}
	\matC_\vecd &=& \matC_\vecx - \matG \matC_\vecz \matG
\end{IEEEeqnarray}
where $\matC_\vecx = \Ex{}{\vecx\vecx^H} \in \opC^{BN \times BN}$ has the following closed-form expression for the case of 1-bit DACs~\cite{van-vleck66a}:
\begin{IEEEeqnarray}{rCl} \label{eq:arcsine_mimo}
	\matC_{\vecx}
	=&&\frac{2 P S }{\pi B N}\bigg(\!\!\arcsin\lefto( \text{diag}(\matC_{\vecz})^{-\frac{1}{2}} \, \Re\{ \matC_{\vecz} \} \, \text{diag}(\matC_{\vecz})^{-\frac{1}{2}}\right) \nonumber\\ 
	&&+j\arcsin\lefto( \text{diag}(\matC_{\vecz})^{-\frac{1}{2}} \, \Im\{ \matC_{\vecz} \} \, \text{diag}(\matC_{\vecz})^{-\frac{1}{2}}\right)\!\!	\bigg).\IEEEeqnarraynumspace 
\end{IEEEeqnarray}
%

%
Through standard manipulations of the mutual information (see, e.g., \cite{jacobsson17d}), we can obtain a lower bound $R_\text{sum}$ on the sum-rate throughput that is explicit in the SINDR~\eqref{eq:sindr_mimo} as follows:
\begin{IEEEeqnarray}{rCl} \label{eq:rate}
	R_\text{sum} &=&  \frac{1}{S}\Ex{\hat\matH}{\sum_{k \in \setS_d}\sum_{u=1}^U \log_2\!\Big( 1 + \sindr_{u,k}(\hat\matH)\Big)}. \IEEEeqnarraynumspace
\end{IEEEeqnarray}
This lower bound corresponds to the rate achievable with a Gaussian codebook and a mismatched nearest-neighbor decoder at the UEs~\cite{lapidoth96b, zhang12a}.

\section{Numerical Results}  \label{sec:numerical}
We focus on a MU-MIMO-OFDM system in which the number of BS antennas is $B=128$ and the number of UEs is $U=16$. We consider a frequency-selective Rayleigh-fading channel with $L=4$ taps and a uniform power delay profile. 
The OFDM numerology is inspired by an LTE system~\cite{3gpp17a}.
Specifically, we assume that the number of occupied subcarriers is $S=300$. The subcarrier spacing is $\Delta{f} = 15$~kHz and the total number of subcarriers (the size of the DFT) is $N = 512$. Hence, the sampling rate of the DACs is $f_s = N \Delta{f} = 7.68~\text{MHz}$ and the OSR is $N / S = 512/300 \approx 1.7$.

\subsection{Spectral Emissions}

\begin{figure}[t!]
\centering
\subfloat[PSD of the transmitted signal.]{\includegraphics[width=\figwidth]{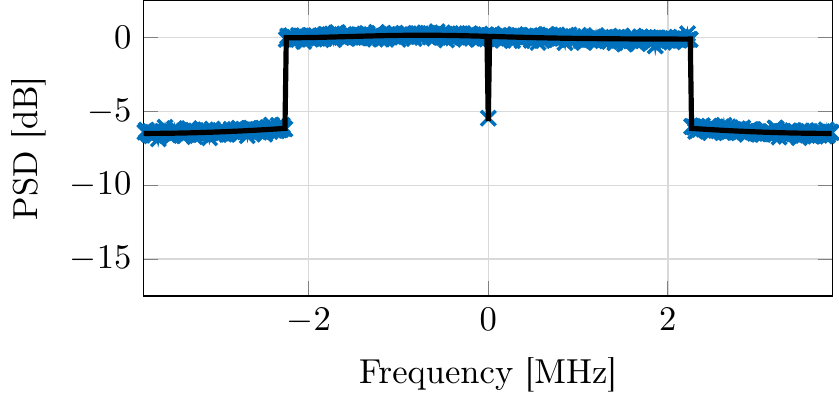}} \\
\subfloat[PSD of the received signal in the noiseless case.]{\includegraphics[width=\figwidth]{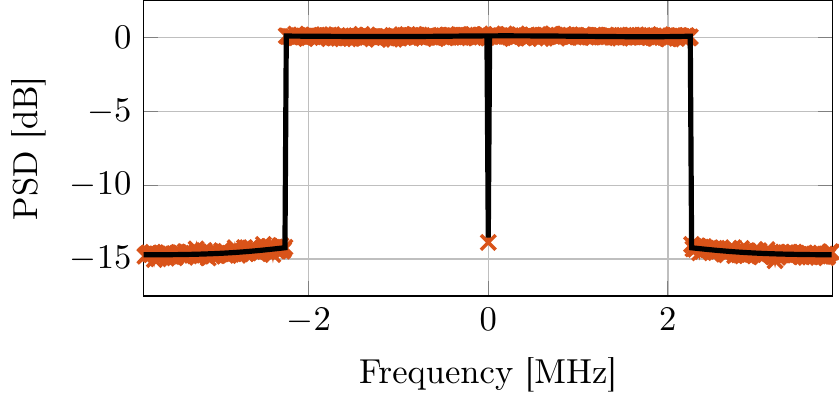}}
\caption{PSD of the transmitted and received signal. The markers correspond to simulated values and the solid lines correspond to analytical results.} 
\label{fig:spectrum}
\end{figure}

In \fref{fig:spectrum}, we plot the normalized PSD of the transmitted signal (averaged over the BS antennas and the channel realizations) and the PSD of the received signal (averaged over the UEs and the channel realizations).
We assume that the BS uses MRT precoding and that the data vector $\vecs$ contains uncoded QPSK symbols.
Furthermore, we assume that the occupied carriers are the first 150 to the left and to the right of the DC carrier (the DC carrier is not occupied). This results in a one-sided bandwidth of $2.25$~MHz (the available one-sided bandwidth is instead $3.84$ MHz).
Numerical simulations are compared with analytic results obtained by computing the PSD of $\vecx$ and of the corresponding output vector using~\eqref{eq:arcsine_mimo}.
We see from the figure that the nonlinearity introduced by the 1-bit-DAC causes severe out-of-band (OOB) distortion at the BS.
Interestingly, however, the OOB distortion of the received signal at the UE is roughly $9$~dB smaller than the OOB distortion of the transmitted signal at the BS. 
This is because the distortion adds up incoherently at the UEs whereas the useful signal is beamformed to the UEs. 
Hence, even though the 1-bit DACs cause significant spectral distortions per antenna, the distortion levels at the UEs are not as severe. 
Nevertheless, the OOB distortion caused by the 1-bit DACs is a significant issue in practical systems as it may cause interference to UEs in adjacent frequency bands.

\subsection{Error-Rate Performance}



\subsubsection{Uncoded BER} 
\begin{figure}[t!]
\centering
 \includegraphics[width=\figwidth]{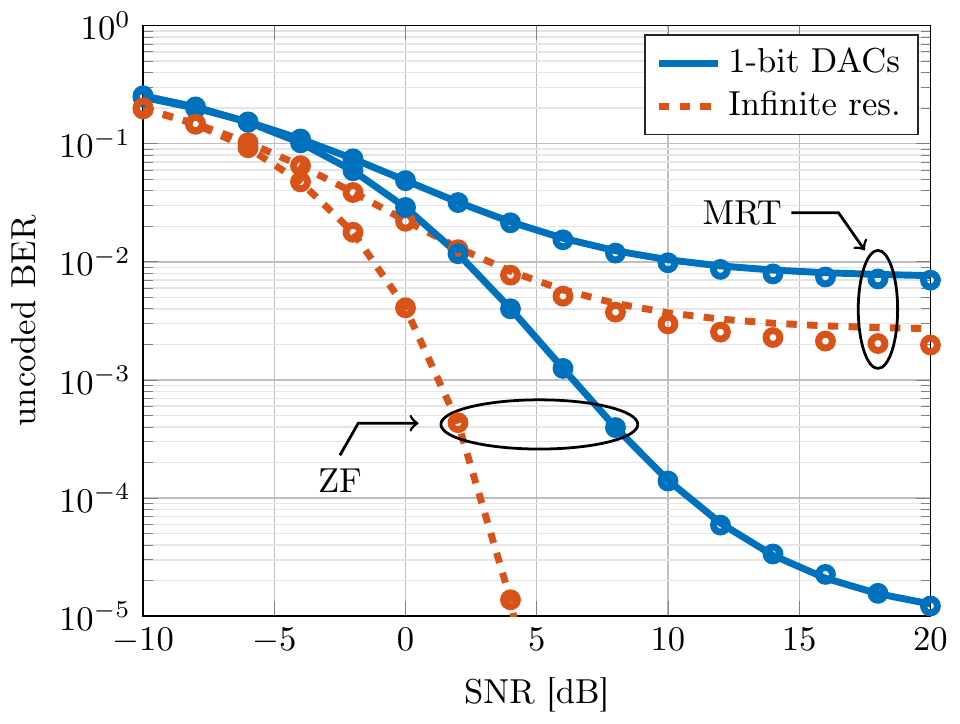}
 \caption{Uncoded BER with QPSK; $U = 16$, $B=128$, $S=300$, $N = 512$. The markers correspond to simulated values and the solid lines correspond to the approximation in~\eqref{eq:ber_approx}. We note that an uncoded BER below $10^{-4}$ is possible with ZF and 1-bit DACs. } 
  \label{fig:ber_tap_uncoded}
\end{figure}

In~\fref{fig:ber_tap_uncoded}, we plot the uncoded BER with QPSK for MRT and ZF precoding as a function of the SNR $P/N_0$. Both the case of 1-bit DACs and of infinite-resolution~DACs are considered. 
The simulated BER values are compared with an analytical approximation for the BER obtained by assuming that the overall noise at the UEs (which includes MU inteference and quantization errors) is Gaussian. 
Under this assumption, the uncoded BER with QPSK can be approximated by
\begin{IEEEeqnarray}{rCl} \label{eq:ber_approx}
	\frac{1}{US}\Ex{\hat\matH}{\sum_{\,u=1}^U \sum_{k \in \setS_d } Q\lefto( \sqrt{\sindr_{u,k}(\hat\matH)}\right)}
\end{IEEEeqnarray}
where $\sindr_{u,k}(\hat\matH)$ is given in \eqref{eq:sindr_mimo}.
We see from \fref{fig:ber_tap_uncoded} that MRT suffers from a relatively high error floor, which is mainly due to residual MU interference, whereas ZF yields better performance, although the gap at $10^{-4}$ from the performance obtainable with infinite-resolution DACs is about $8$~dB.
%
The approximation~\eqref{eq:ber_approx} is accurate over the entire range of SNR values considered in the figure. 
This suggests that the overall noise per subcarrier at the UEs can indeed be modeled accurately as a Gaussian random variable.

\begin{figure}[t!]
\centering
 \includegraphics[width=\figwidth]{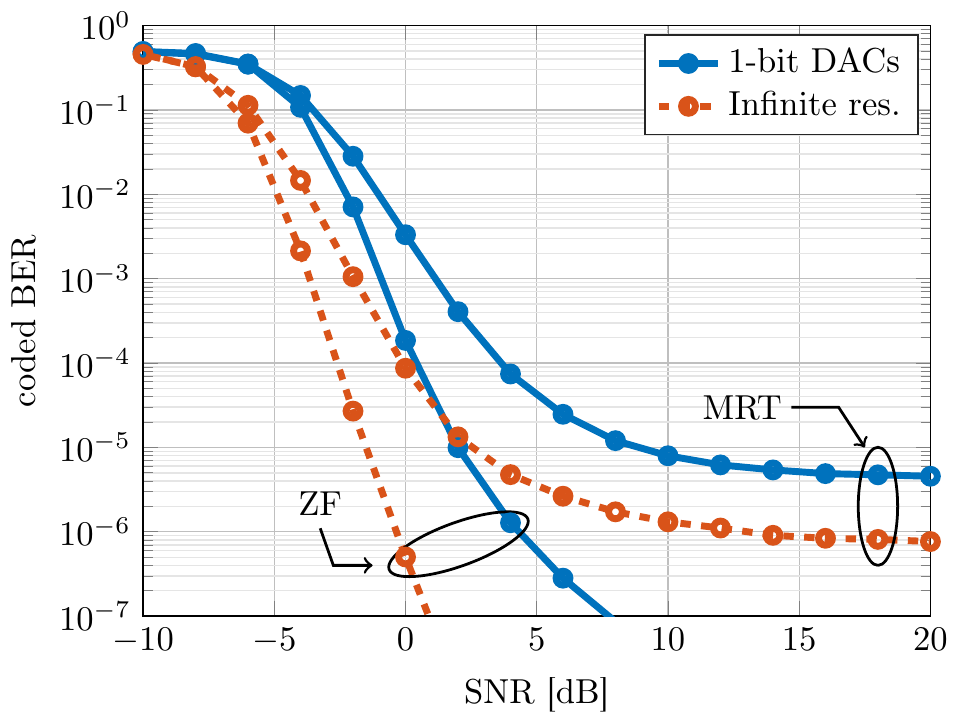}
 \caption{Coded BER with QPSK for a rate-1/2 convolutional code, and with hard-decision decoding at the UEs; $U = 16$, $B=128$, $S=300$, $N = 512$. Low coded BERs are achievable with ZF and 1-bit DACs.} 
  \label{fig:ber_tap_coded}
\end{figure}

\subsubsection{Coded BER} In~\fref{fig:ber_tap_coded} we show the coded BER for MRT and ZF, for the case of QPSK modulation and rate-$1/2$ convolutional code. 
The length of the codewords is~$6000$~bits and the QPSK symbols are randomly interleaved over $10$~consecutive OFDM symbols. 
The UEs perform symbol-wise nearest-neighbor decoding and hard-input Viterbi decoding (extension to soft-input decoding is part of ongoing work).
With ZF, the gap to the infinite-resolution performance is $4.5$\,dB for a target BER of $10^{-6}$. 
As in the uncoded case, MRT is limited by MU interference; the error floor is, however, below $10^{-5}$.

\subsection{Achievable Rate}

\begin{figure}[t!]
\centering	
\includegraphics[width=\figwidth]{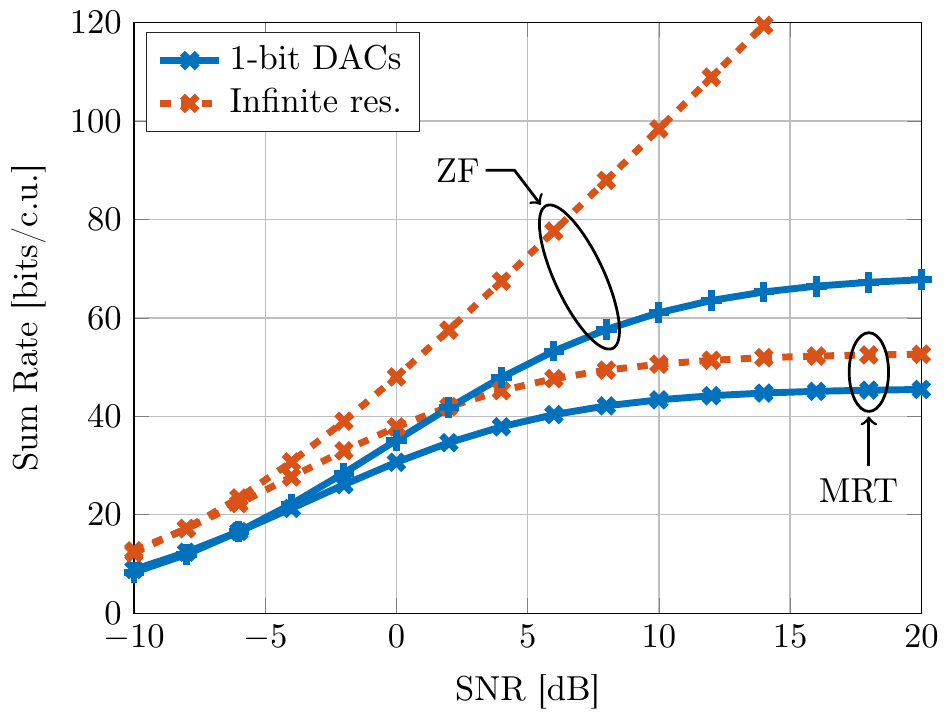}
\caption{Achievable sum-rate with Gaussian inputs; $U = 16$, $B=128$, $S=300$, $N = 512$. High sum rates are achievable with ZF and~1-bit~DACs.} 
\label{fig:rate}
\end{figure}

In~\fref{fig:rate}, we show the sum-rate throughput~\eqref{eq:rate} achievable with Gaussian inputs and mismatched nearest-neighbor decoding as a function of the SNR.
We observe that, similarly to the frequency-flat case~\cite{jacobsson17d}, a high sum-rate throughput can be achieved. 
Specifically, a sum-rate throughput exceeding $64$-bits per channel use (corresponding to $4$~bits per channel use per UE) can be achieved for SNR values beyond $13$\,dB. 

\subsection{Impact of Oversampling} \label{sec:oversampling}

\begin{figure}[t!]
\centering	
\includegraphics[width=\figwidth]{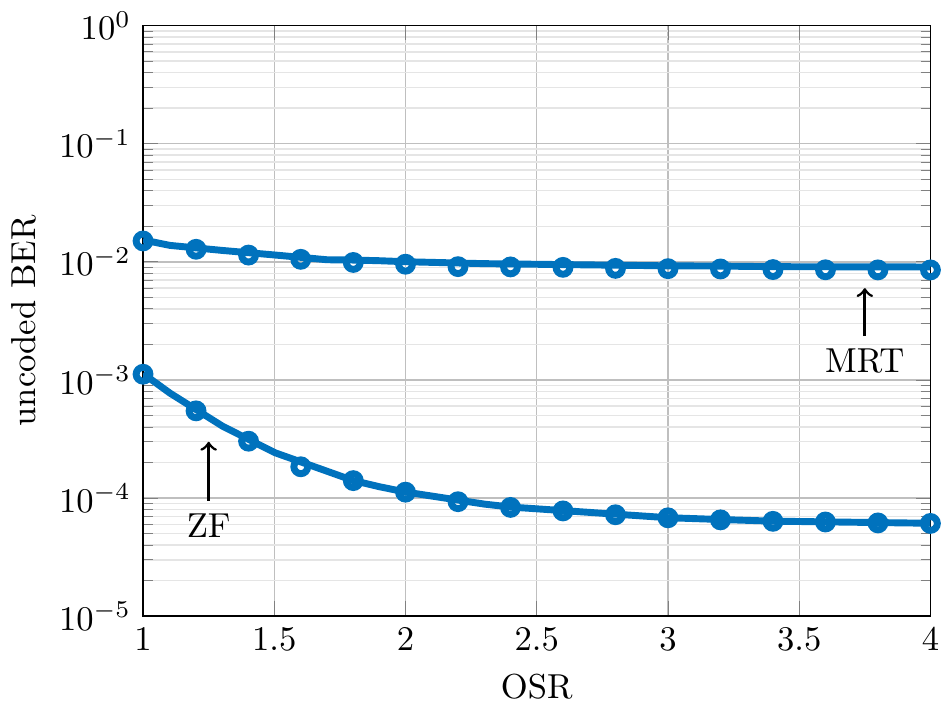}
\caption{Impact of the OSR on the uncoded BER; SNR$=10$~dB, $U = 16$, $B=128$, $S=300$. Increasing the OSR improves the uncoded BER.} 
\label{fig:osr}
\end{figure}
%
In~\fref{fig:osr}, we investigate the impact of the OSR on the uncoded BER.
Specifically, we plot the uncoded BER for the case of uncoded QPSK as a function of the OSR for both MRT and ZF. 
The SNR, defined as $P/N_0$, is set to $10$~dB. 
We note that, for ZF, the uncoded BER can be considerably improved by operating the DACs at a sampling rate higher than the symbol rate. 
Indeed, the uncoded BER with ZF can be decreased by an order of magnitude  compared to the symbol-rate sampling case (OSR$=1$)  by operating the DACs at twice the symbol rate. However, further increasing the  sampling rate yields only marginal performance~gains.

\section{Conclusions} \label{sec:conclusions}

We have characterized the performance in terms of BER and achievable rate of a MU-MIMO-OFDM downlink system, where the BS is equipped with 1-bit DACs and uses linear precoding. 
Using Bussgang's theorem, we have obtained a closed-form expression for the SINDR, which we have then used to obtain a lower bound on the achievable sum-rate with Gaussian inputs and an accurate approximation for the uncoded BER with QPSK.
Through numerical simulations, we have illustrated that low uncoded and coded BERs and high sum-rates are achievable despite the severe nonlinearity introduced by the 1-bit DACs, at the cost of unwanted OOB emissions. 

Extensions of our analysis to the case of multi-bit DACs and to nonlinear precoders is part of ongoing work. Developing methods for reducing OOB emissions is also part of ongoing~work. 

\bibliographystyle{IEEEtran}
\begin{spacing}{.95}
\bibliography{IEEEabrv,confs-jrnls,publishers,svenbib}	

\begin{thebibliography}{10}
\providecommand{\url}[1]{#1}
\csname url@samestyle\endcsname
\providecommand{\newblock}{\relax}
\providecommand{\bibinfo}[2]{#2}
\providecommand{\BIBentrySTDinterwordspacing}{\spaceskip=0pt\relax}
\providecommand{\BIBentryALTinterwordstretchfactor}{4}
\providecommand{\BIBentryALTinterwordspacing}{\spaceskip=\fontdimen2\font plus
\BIBentryALTinterwordstretchfactor\fontdimen3\font minus
  \fontdimen4\font\relax}
\providecommand{\BIBforeignlanguage}[2]{{%
\expandafter\ifx\csname l@#1\endcsname\relax
\typeout{** WARNING: IEEEtran.bst: No hyphenation pattern has been}%
\typeout{** loaded for the language `#1'. Using the pattern for}%
\typeout{** the default language instead.}%
\else
\language=\csname l@#1\endcsname
\fi
#2}}
\providecommand{\BIBdecl}{\relax}
\BIBdecl

\bibitem{rusek14a}
F.~Rusek, D.~Persson, B.~Kiong, E.~G. Larsson, T.~L. Marzetta, O.~Edfors, and
  F.~Tufvesson, ``Scaling up {MIMO}: Opportunities and challenges with very
  large large arrays,'' \emph{{IEEE} Signal Process. Mag.}, vol.~30, no.~1, pp.
  40--60, Jan. 2013.

\bibitem{larsson14a}
E.~G. Larsson, F.~Tufvesson, O.~Edfors, and T.~L. Marzetta, ``Massive {MIMO}
  for next generation wireless systems,'' \emph{{IEEE} Commun. Mag.}, vol.~52,
  no.~2, pp. 186--195, Feb. 2014.

\bibitem{wen15b}
C.-K. Wen, C.-J. Wang, S.~Jin, K.-K. Wong, and P.~Ting, ``{B}ayes-optimal joint
  channel-and-data estimation for massive {MIMO} with low-precision {ADC}s,''
  \emph{{IEEE} Trans. Signal Process.}, vol.~64, no.~10, pp. 2541--2556, Jul.
  2015.

\bibitem{mollen16c}
C.~Moll{\'e}n, J.~Choi, E.~G. Larsson, and R.~W. {Heath Jr.}, ``Uplink
  performance of wideband massive {MIMO} with one-bit {ADCs},'' \emph{{IEEE}
  Trans. Wireless Commun.}, vol.~16, no.~1, pp. 87--100, Oct. 2016.

\bibitem{jacobsson17b}
S.~Jacobsson, G.~Durisi, M.~Coldrey, U.~Gustavsson, and C.~Studer, ``Throughput
  analysis of massive {MIMO} uplink with low-resolution {ADCs},'' \emph{{IEEE}
  Trans. Wireless Commun.}, vol.~16, no.~6, pp. 4038--4051, Jun. 2017.

\bibitem{li17b}
Y.~Li, C.~Tao, G.~Seco-Granados, A.~Mezghani, A.~L. Swindlehurst, and L.~Liu,
  ``Channel estimation and performance analysis of one-bit massive {MIMO}
  systems,'' \emph{{IEEE} Trans. Signal Process.}, vol.~65, no.~15, pp.
  4075--4089, Aug. 2017.

\bibitem{saxena16b}
A.~K. Saxena, I.~Fijalkow, and A.~L. Swindlehurst, ``Analysis of one-bit
  quantized precoding for the multiuser massive {MIMO} downlink,'' \emph{{IEEE}
  Trans. Signal Process.}, vol.~65, no.~17, pp. 4624--4634, Sep. 2017.

\bibitem{li17a}
Y.~Li, C.~Tao, A.~L. Swindlehurst, A.~Mezghani, and L.~Liu, ``Downlink
  achievable rate analysis in massive {MIMO} systems with one-bit {DACs},''
  \emph{{IEEE} Commun. Lett.}, vol.~21, no.~7, pp. 1669--1672, Jul. 2017.

\bibitem{jacobsson17d}
S.~Jacobsson, G.~Durisi, M.~Coldrey, T.~Goldstein, and C.~Studer, ``Quantized
  precoding for massive {MU-MIMO},'' \emph{{IEEE} Trans. Commun.}, 2017, to
  appear.

\bibitem{jacobsson16d}
------, ``Nonlinear 1-bit precoding for massive {MU-MIMO} with higher-order
  modulation,'' in \emph{Proc. Asilomar Conf. Signals, Syst., Comput.}, Pacific
  Grove, CA, USA, Nov. 2016, pp. 763--767.

\bibitem{castaneda17a}
O.~Casta\~{n}eda, S.~Jacobsson, G.~Durisi, M.~Coldrey, T.~Goldstein, and
  C.~Studer, ``1-bit massive {MU}-{MIMO} precoding in {VLSI},'' \emph{IEEE J.
  Emerging Sel. Topics Circuits Syst.}, 2017, to appear.

\bibitem{jedda16a}
H.~Jedda, J.~A. Nossek, and A.~Mezghani, ``Minimum {BER} precoding in 1-bit
  massive {MIMO} systems,'' in \emph{{IEEE} Sensor Array and Multichannel
  Signal Process. Workshop (SAM)}, Rio de Janeiro, Brazil, Jul. 2016.

\bibitem{jedda16b}
H.~Jedda, A.~Mezghani, J.~Munir, F.~Steiner, and J.~A. Nossek, ``Spatial coding
  based on minimum {BER} in 1-bit massive {MIMO} systems,'' in \emph{Proc.
  Asilomar Conf. Signals, Syst., Comput.}, Pacific Grove, CA, USA, Nov. 2016,
  pp. 753--757.

\bibitem{tirkkonen17a}
O.~Tirkkonen and C.~Studer, ``Subset-codebook precoding for 1-bit massive
  multiuser {MIMO},'' in \emph{Conf. Inf. Sciences Syst. (CISS)}, Baltimore,
  MD, USA, Mar. 2017.

\bibitem{guerreiro16a}
R.~D. J.~Guerreiro and P.~Montezuma, ``Use of 1-bit digital-to-analogue
  converters in massive {MIMO} systems,'' \emph{{IEEE} Electron. Lett.},
  vol.~52, no.~9, pp. 778--779, Apr. 2016.

\bibitem{bussgang52a}
J.~J. Bussgang, ``Crosscorrelation functions of amplitude-distorted {Gaussian}
  signals,'' Res. Lab. Elec., Cambridge, MA, Tech. Rep. 216, Mar. 1952.

\bibitem{rowe82a}
H.~E. Rowe, ``Memoryless nonlinearities with {Gaussian} inputs: Elementary
  results,'' \emph{Bell Labs Tech.~J.}, vol.~61, no.~7, pp. 1519--1525, Sep.
  1982.

\bibitem{van-vleck66a}
J.~H. Van~Vleck and D.~Middleton, ``The spectrum of clipped noise,''
  \emph{Proc. {IEEE}}, vol.~54, no.~1, pp. 2--19, Jan. 1966.

\bibitem{lapidoth96b}
A.~Lapidoth, ``Nearest neighbor decoding for additive non-{Gaussian} noise
  channels,'' \emph{{IEEE} Trans. Inf. Theory}, vol.~42, no.~5, pp. 1520--1529,
  Sep. 1996.

\bibitem{zhang12a}
W.~Zhang, ``A general framework for transmission with transceiver distortion
  and some applications,'' \emph{{IEEE} Trans. Commun.}, vol.~60, no.~2, pp.
  384--399, Feb. 2012.

\bibitem{3gpp17a}
3GPP, ``{LTE}; evolved universal terrestrial radio access ({E-UTRA}); user
  equipment ({UE}) radio transmission and reception,'' Apr. 2017, {TS} 36.101
  version 14.3.0 Rel.~14.

\end{thebibliography}
\end{spacing}

\end{document}